\documentclass[12pt]{article}

\usepackage{graphicx}
\usepackage{amsmath,amsthm}
\usepackage{amssymb,latexsym}
\usepackage{enumerate}
\usepackage[shortlabels]{enumitem}
\usepackage{epigraph}
\usepackage[authoryear]{natbib}
\usepackage{mathrsfs}
\usepackage{verbatim}
\usepackage{mathabx}
\usepackage{relsize}
\usepackage{mathtools}
\usepackage{textcomp}
\usepackage{accents}
\usepackage{bm}
\usepackage[titletoc,title]{appendix}
\usepackage{url}
\usepackage{setspace}
\usepackage{textcomp}
\usepackage{pstricks}
\usepackage{wasysym}
\makeatletter
\g@addto@macro\bfseries{\boldmath}
\makeatother

\theoremstyle{plain}

\theoremstyle{definition}

\theoremstyle{remark}

\newtheorem{exmp}{Example}[section]

\usepackage{geometry}
\usepackage{doc}
\usepackage{epstopdf}
\begin{document}

% Add the title section.
\begin{titlepage}
  \centering
  {\scshape\Large {Does it Pay to Buy the Pot in the Canadian 6/49 Lotto? Implications for Lottery Design}\par}
  \vspace{1.5cm}
  {\scshape\Large June 7, 2017\par}
  \vspace{2cm}
  {\Large Steven D. Moffitt\textsuperscript{\textdagger} and William T. Ziemba\textsuperscript{\textdaggerdbl}\par}
  \vfill
  \textdagger Adjunct Professor of Finance, Stuart School of Business, Illinois Institute of Technology and 
              Principal, Market Pattern Research, Inc.\par
  \textdaggerdbl Alumni Professor of Financial Modeling and Stochastic Optimization (Emeritus),
  University of British Columbia, Vancouver, BC, and Distinguished Visiting Research Associate, Systemic Risk Centre,
  London School of Economics, UK
\end{titlepage}

% Add an abstract.
\abstract{
\noindent Despite its unusual payout structure, the Canadian 6/49 Lotto\textcopyright is one of the few government sponsored lotteries that has the potential for a favorable strategy we call ``buying the pot.'' By ``buying the pot'' we mean that a syndicate buys each ticket in the lottery, ensuring that it holds a jackpot winner. We assume that the other bettors independently buy small numbers of tickets. This paper presents (1) a formula for the syndicate's expected return, (2) conditions under which buying the pot produces a significant positive expected return, and (3) the implications of these findings for lottery design.
}

%************************************************************************************************
%************************************************************************************************
\section[Introduction]{Introduction}

\cite{moffitt:ziemba:2017a} show that expected returns of $10\%$-$25\%$ can be achieved under certain conditions from betting all the tickets in a lottery that pays its entire jackpot in equal shares to winning ticket holders. For many large government lotteries, this strategy of ``buying the pot'' is not feasible because the logistical problems are insurmountable. In the California Powerball Lottery\textcopyright, for example, the number of ticket combinations is over $175,000,000$ and the rules do not allow betting large numbers of combinations on single paper tickets. 

The Canadian 6/49 Lotto\textcopyright , however, has a large but manageable number of ticket combinations ($13,983,816$) and allows paper tickets that have multiple combinations. The 6/49 is played in other countries, including the UK. Here we focus on the Canadian version. 

The purpose of this paper is threefold: (1) to modify the \emph{pure jackpot model} in \cite{moffitt:ziemba:2017a} to accomodate the irregular payout features of the Canadian 6/49 Lotto, (2) to derive conditions under which the expected return from buying the pot is positive, and (3) to discuss the implications of our findings for lottery design.

%************************************************************************************************
%************************************************************************************************
\section{Previous Work and Instances of Buying the Pot}\label{S:PreviousWorkAndInstancesOfBuyingThePot}

%The results in this paper are mostly applications to the Canadian 6/49 Lotto of a \emph{pure jackpot model} in \cite{moffitt:ziemba:2017a}. 
Each lottery has the following rules --- players buy tickets and the winning ticket is selected using an equiprobable drawing. Those who hold the winning ticket share equally in a jackpot that consists of a carryover pot from the previous lottery plus an after tax portion the monies wagered. If there is no winner, the jackpot pool carries over to the next drawing. There can be multiple carryovers.

\cite{moffitt:ziemba:2017a} use the following assumptions and notation to analyze the \emph{pure jackpot model}:
\begin{itemize}
\item Each lottery has $t$ tickets costing $\$1$ apiece.
\item A single winning ticket $w$, $1 \le w \le t$ is drawn from $i=1,\ldots,t$ using probabilities $p_i = 1/t$.
\item The syndicate buys one of each ticket for a total of $t$ tickets, and $c$ individuals (the ``crowd'') independently buy one ticket apiece using probabilities $q_i$, $1 \le i \le t$.
\item A cash jackpot $v = a + (t + c)(1 - x)$ is awarded in equal shares to all holders of the winning ticket $w$, where $a \ge 0$ is the current carryover from the previous lottery draws, $c$ is the number of tickets bet by the crowd, and $x$ is the the (fractional) take. 
\end{itemize}

\noindent \cite{moffitt:ziemba:2017a} show the following for the pure jackpot model:
\begin{enumerate}[label=\bfseries {(\arabic*)},ref={\thesubsection (\arabic*)}]

  \item{\textit{Recursion:}} When $t$ and $c$ are large, $q_i = 1/t$ for each $i$, and $X$ is the random number of winning tickets held by the crowd, the expected value $E\left[ \frac{n}{n+X} \right]$, $n$ an integer $\ge 1$, is to close approximation equal to
    \begin{equation} \label{E:EVrecursion}
      E\left[ \frac{n}{n+X} \right] = \begin{cases}
                                        \frac{1}{\lambda(c)} \left( 1 - e^{-\lambda(c)} \right) & n=1 \\
                                        \frac{n}{\lambda(c)} \left\{ 1 -  E\left[ \frac{n-1}{n-1+X} \right] \right\} & n > 1,
                                      \end{cases}
    \end{equation}
    where $\lambda(c) = c/t$. \label{Enum:Recursion}

  \item{\textit{Condition under which Buying the Pot has Positive Expected Return:}} The expected gain for a syndicate that bets one of each ticket is positive
    \begin{equation}
      (a + (t + c)(1 - x)) \, E\left[ \frac{1}{1+X} \right] \, - \, t \, > 0 \label{E:FairSplitCondition}
    \end{equation}
    provided that $a/(t + c) - x \ge 0$. Since $a/(t + c) - x$ is the after tax value of a ticket assuming the pot $a$ is fairly split, this condition implies that a syndicate earns more than a fair split of the jackpot. In a lottery with no take, the returns to the syndicate typically range between 10\% and 25\%. \label{Enum:BTPEdge}

  \item{\textit{Optimal Strategies:}} 
    \begin{enumerate}
      \item The best returning strategy for the crowd consists of using $q_i = 1/t$ for each $i$. 
      \item Let $E_q[X/(1+X)]$ be the expectation for a crowd that bets with probability vector $q = (q_1, \ldots, q_t)'$, and let $1_t/t$ be the probability t-vector that has $1/t$ for each entry. Then if $q \ne 1_t/t$
      \begin{equation}
        E_q[X/(1+X)] < E_{1_t/t}[X/(1+X)]. \label{E:OptimalityOfCrowdProportionalBetting}
      \end{equation}
  \end{enumerate} \label{Enum:OptimalStrategy}
\end{enumerate}

Several studies of lottery strategy and design have appeared in the economic literature. \cite{chernoff1980analysis,Chernoff1981} studied the Massachusetts Numbers Game, proposing that playing unpopular numbers might be a winning strategy. However, the results from a test were disappointing because of learning (unpopular numbers became less unpopular) and gambler's ruin (betting funds were exhausted). \cite{ziemba1986dr} carry this further and study various Canadian lotto games, their unpopular numbers and the uniformity of betting. \cite{10.2307/2290349} has additional discussion of this latter point and \cite{10.2307/3314913}, \cite{10.2307/2290073} and  \cite{Ziemba2008183} further analyze unpopular numbers. \cite{citeulike:1337256} review the behavioral evidence in efficient markets for a persistence of betting at unfavorable odds. \cite{RePEc:inm:ormnsc:v:38:y:1992:i:11:p:1562-1585} investigate the use of Kelly optimal wagering on unpopular tickets and find that this strategy has positive expectation, but the waiting time to achieve reliable gains with high probability is millions of years! \cite{10.2307/1942724} discuss behavioral bases of betting and along with \cite{Walker2008459}, discuss design considerations for lotteries. None of these studies consider the strategy that in a short time achieves reliable gains --- buying the pot.

There are anecdotal accounts of successful buyings of the pot. One putative attempt involved a syndicate that tried but failed to buy all tickets. But they were lucky, having had time to bet only about $70\%$ of all tickets according to one source and $85\%$ according to another (\cite{NYTimes:US:BuyThePot1992}). The syndicate ostensibly bet about \$5 million and won about \$27 million. 

There are examples of when it was optimal to buy the pot or betting was advantageous. In June 1984 four western Canadian provinces jointly ran the \emph{Lotto West 6/8/56}, in which players choose six numbers from a field of 56, but eight winning numbers and a bonus number are drawn. The jackpot is shared among all tickets that select six of the eight drawn, second price among all that had five of six, and other payouts to those having five of six plus the bonus, four of six or three of six. These rules make the jackpot about twelve times easier to hit than the 6/49 Lotto: 1 in $1,159,587$ (See \cite{ziemba1986dr}). 

In 1987, the provinces went their own ways, at which time the BC Lotto Corporation had about \$10 million in unclaimed prize money. Rather than donate it, they created a version of Lotto 6/8/56 to give it back on March 27, 1987. As before eight numbers were drawn from 56, but players could now choose 1, 2, 3, 4, 5 or 6 numbers on a ticket. A schedule of payouts was published for 1/1, 2/2, /3/3, 4/4, 5/5 and for 3/6, 4/6, 5/6 and 6/6. With these payouts, the expected payback on a \$1 ticket was $\$0.385$. To promote the game, the Corporation offered six tickets for the price of one, for an expected return of \$0.385 times 6, or \$2.31, a $131\%$ edge (\cite{drz:ColumnsOnRacing}). Ziemba and colleagues at the University of British Colombia knew that individual tickets had a positive expected return, and in a makeshift effort, they bought about 13,000 of the combinations. They made a nice return, but spent hours buying and then locating the winning tickets. Some $\$3.5$ million was paid out of the unclaimed prize fund.

A game where it was optimal to buy the pot was the 5/40 Lotto played in British Colombia and Rhode Island. Ninety-one percent of the net pool went to 5/5 with a minimum shared pool of $\$150,000$ and maximum of $\$300,000$. There were small prizes for 1+, 2+, 3, 4 and 4+ where ``+'' means getting the sixth bonus number correct. There were $658,008$ combinations. But the jackpot that had built up slowly fell because the public viewed it as unwinnable, so it became a prime target for buying the pot; see \cite{drz:ColumnsOnRacing}.

%************************************************************************************************
%************************************************************************************************
\section[Rules of the 6/49 Lotto]{Rules of the 6/49 Lotto}\label{S:RulesOfThe649Lotto}

A \emph{ticket} in the 6/49 Lotto is a unique choice of $6$ different numbers from integers $1$ to $49$. Thus the total number of tickets is the number of combinations of $49$ things taken $6$ at a time:
\begin{equation}
  t = \binom{49}{6} = \frac{49!}{43! 6!} = 13, 983, 816. \label{Nbr-6/49-Tickets}
\end{equation}

The Canadian 6/49 Lotto holds drawings twice a week and lumps together the monies wagered for purposes of awarding prizes, whose allocation is described below. On the drawing day, 6 numbers (the ``\emph{winning numbers}'') are selected equiprobably and without replacement from 1, 2, \ldots 49. Following that, a $7^{th}$ ``\emph{bonus number}'' is selected. 

We introduce notation to describe types of prize-wining tickets. A $x$/6- ticket is one that contains exactly $x$ of the six winning numbers but does not contain the bonus number and a $x$/6+ ticket is one that contains exactly $x$ of the 6 numbers plus the bonus number. A x/6 ticket contains x of the 6 numbers, irrespective of the status of the bonus number; it is therefore a union of types x/6- and x/6+. A 5/6-, for example, contains exactly 5 of the 6 winning numbers with the other not being the bonus number, and a 5/6+ ticket contains exactly 5 of the 6 winning numbers plus the bonus number. For example, if the six numbers drawn were 46, 13, 4, 21, 38, 25 and the bonus number was 43 then ticket 1-4-20-21-32-43 would be a 2/6+ ticket because it contains 4 and 21 from the six plus the bonus number. Similarly, ticket 4-13-21-25-43-46 would be a 5/6+ ticket. 

\subsection{Rules for the Original Lottery: 1982-2004}

Table \ref{Ta:PrizeAllocIn649:1982-2004} has the initial 6/49 payout scheme (1982-2004) for 3/6, 4/6, 5/6-, 5/6+ and the \emph{Jackpot} 6/6. The cost of a single ticket was \$1, with the lottery sponsors taking 55\% of each daily \emph{betting pool} and committing the remaining 45\% (the ``\emph{prize pool}'') for player payouts. The 45\% \emph{prize pool} was allocated as follows: all 3/6 tickets were paid \$10, and the remainder was paid to holders of 4/6, 5/6-, 5/6+ and 6/6 using percentage allocation rules in Table \ref{Ta:PrizeAllocIn649:1982-2004}. That game is analyzed thoroughly in \cite{ziemba1986dr}. See also \cite{10.2307/2290073}. For other analyses of such games, see \cite{citeulike:1337256} and  \cite{Haigh2008481}.

\scriptsize
\begin{table}[ht!]
  \begin{center}
    \caption{\small{Allocation of Prizes in the 6/49 Pools Fund: 1982-2004.}}
    \label{Ta:PrizeAllocIn649:1982-2004}
    \scriptsize
    \begin{tabular} {lrlll}
      \rule[-4pt]{0pt}{10pt} \\
      \hline
      \rule[-4pt]{0pt}{10pt} \\
      Prize & Combinations & Probability & Allocation Rule & Type \\
      \rule{0pt}{2pt} \\
      \hline
      \rule{0pt}{0pt} \\
      $6/6$  &          $1$ & $p_1 \sim 7.1e$-$08$ &  50\% of the Pools Fund       & Share \\
      $5/6+$ &          $6$ & $p_2 \sim 4.3e$-$07$ &  15\% of the Pools Fund       & Share \\
      $5/6-$ &        $252$ & $p_3 \sim 1.8e$-$05$ &  12\% of the Pools Fund       & Share \\
      $4/6$  &     $13,545$ & $p_4  = 0.000969$    &  23\% of the Pools Fund       & Share \\
      $3/6$  &    $246,820$ & $p_5  = 0.017650$    &  $\$10$ per ticket            & Fixed \\
      No Win & $13,723,192$ & $p_8  = 0.981362$    &  Non-winner = \$0             & Returns 0 \\
      \rule{0pt}{0pt} \\
      \hline
    \end{tabular}
  \end{center}
\end{table}
\normalsize

\begin{figure}[ht!]
  \centering
    \includegraphics[scale=0.80]{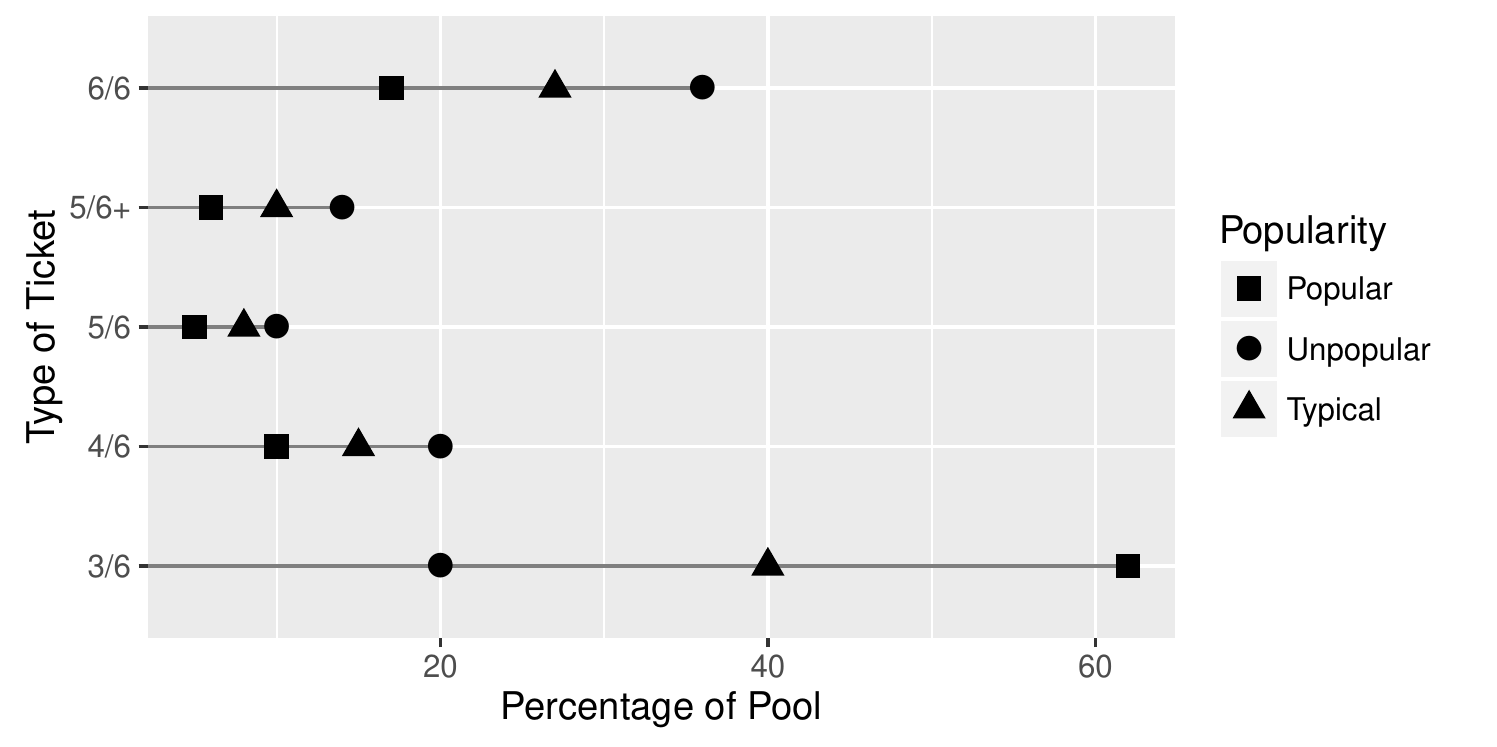}
    \caption{\small{Prize shares for 3/6, 4/6, 5/6, 5/6+ and 6/6 in the original 6/49 Lotto when the drawing has Popular Numbers ($\blacksquare$), 
                    Unpopular Numbers ({\Large \textbullet}) and Average Numbers ($\blacktriangle$). For popular numbers, there is a  
                    large increase in the payouts for 3/6 tickets ($\bm{>}$ 60\%) at the expense of other winning outcomes. Source: \cite{ziemba1986dr}.}}
    \label{G:PoolShare}
\end{figure}

Figure \ref{G:PoolShare} shows the leveraging effect of the fixed \$10 3/6 prize when there are average numbers, popular numbers, and unpopular numbers. The impact of popular vs. unpopular numbers selected in the drawing is significant, producing a 17\% versus a 36\% jackpot share. The large prizes 5/6-, 5/6+ and 6/6 for unpopular numbers in the drawing are typically seven times larger than for popular ones. See examples in \cite{ziemba1986dr}.
\normalsize

\subsection{Rules for the Current Lottery: 9/18/2013 - }

In the 6/49 Lotto, new rules were introduced in June, 2004 and again on September 18, 2013. We discuss only the latter rules. These included (1) a single ticket cost of \$3, (2) three fixed prizes, the same 3/6 paying \$10, a 2/6+ paying \$5 and a 2/6- that earns a free play at the next drawing, and (3) altered payout percentages for 4/6, 5/6-, 5/6+ and 6/6 (Table \ref{Ta:PrizeAllocIn6492013-}), (4) an increase in the take from 55\% to 60\%, and (5) a greater allocation to 6/6 winners. The intention of these changes was to increase sales by growing jackpots faster, and creating of many small consolation prices (2/6-, 2/6+ and 3/6). This is a typical convex prize structure where most of the daily payout goes to the smallest (to make them feel that the lottery is winnable) and to the largest (to show that a huge gain can be made). Ziemba has used this in lottery consulting over the years. \cite{RePEc:eee:jfinec:v:13:y:1984:i:2:p:253-282} call this a ``silver lining'' for non-winners.

We call the number of tickets bet at a drawing (twice a week in the 6/49), the \emph{ticket pool}, contributors to which are the crowd in amount $c$ and the syndicate in amount $t$. Thus the total number of tickets bet is $c + t$. The \emph{betting pool} $d_{\!_{BP}}$ is the total number of dollars contributed by the bettors. The \emph{betting pool} is divided among the lottery sponsors and the bettors as follows: 
\begin{description}

\item{\it Sponsors.} Sponsors (the state, the lottery organization) receive $0.60*d_{\!_{BP}}$, with the remaining $0.40*d_{\!_{BP}}$, the \emph{prize pool}, awarded as prizes or added to the carryover pool as indicated below. The ``lottery take'' $0.60*d_{\!_{BP}}$ is used to cover expenses of running the lottery and to provide funds for community and government services and for donations. The lottery itself, however, is run by a non-governmental company.

\item{\it Prize Distribution.} The \emph{prize pool} has eight classes ($i = 1, 2, \ldots, 8$) of payouts grouped into four types: (A) fixed dollar (2/6+ and 3/6), (B) free play in the next lottery (2/6-), (C) payouts that split among 4/6, 5/6-, 5/6+ and 6/6 tickets the remaining \emph{prize pool} after deductions for type (A) and (B) payouts, and (D) non-winner tickets that receive no payout. 

Table \ref{Ta:PrizeAllocIn6492013-} details these payouts by showing in the first column the type of ticket, in the second column a notation for the number of each class determined after the random, equiprobable drawing of 6 numbers and a bonus, the third column showing the notation for the class, the fourth column the number of tickets matching a randomly drawn ticket, the fifth column having the probability that a randomly chosen ticket is in the class, the sixth column having the allocation rule and the last, whether the ticket payout is fixed, shared as part of a pool, or returns nothing. 

The 2/6+ and 3/6 tickets receive \$5 and \$10, respectively, and  2/6- tickets receive a free play in the next lottery, but a charge of \$1.41 is applied to the \emph{prize pool}. See Example \ref{Exmp:ExampleOfPrizePayouts} for details. The payouts are shown in the first four lines of the table for 4/6, 5/6-, 5/6+ and 6/6 tickets. These type (C) tickets share the remainder of the $0.40*d_{\!_{BP}}$ after deductions for tickets of types (A) and (B). The amount $0.40*d_{\!_{BP}} - (\text{payouts to 2/6+, 3/6 and charges for 2/6-})$ is called the \emph{Pools Fund}. Type (C) tickets share in a pool whose percentage of the total bets varies greatly, depending on the winning numbers of 2/6+, 3/6 and free plays. The lottery also guarantees a \$5,000,000 pool to holders of 6/6 tickets.

\scriptsize
\begin{table}[ht]
  \begin{center}
    \caption{\small{Allocation of Prizes in the Current 6/49 Pools Fund.}}
    \label{Ta:PrizeAllocIn6492013-}
    \scriptsize
    \begin{tabular} {lclrlll}
      \rule[-4pt]{0pt}{10pt} \\
      \hline
      \rule[-4pt]{0pt}{10pt} \\
           & \# Crowd   &       & \# Combinations &             & Allocatio & Share  \\
      Type &   Tickets  & Class & for any ticket  & Probability &  Rule     & Status \\
      \rule{0pt}{2pt} \\
      \hline
      \rule{0pt}{0pt} \\
      (C) & $N_1$ & $6/6$   &          $1$ & $p_1 \sim 7.1e$-$08$ &  79.5\% of the Pools Fund        & Share \\
      (C) & $N_2$ & $5/6+$  &          $6$ & $p_2 \sim 4.3e$-$07$ &  6\% of the Pools Fund           & Share \\
      (C) & $N_3$ & $5/6-$  &        $252$ & $p_3 \sim 1.8e$-$05$ &  5\% of the Pools Fund           & Share \\
      (C) & $N_4$ & $4/6$   &     $13,545$ & $p_4  = 0.000969$    &  9.5\% of the Pools Fund         & Share \\
      (A) & $N_5$ & $3/6$   &    $246,820$ & $p_5  = 0.017650$    &  $\$10$ per ticket               & Fixed \\
      (A) & $N_6$ & $2/6+$  &    $172,200$ & $p_6  = 0.012314$    &  $\$5$ per ticket                & Fixed \\
      (B) & $N_7$ & $2/6-$  &  $1,678,950$ & $p_7  = 0.120064$    &  free play ($\$1.41$ deduction)  & Fixed \\
      (D) & $N_8$ & No Win  & $11,872,042$ & $p_8  = 0.848984$    &  Non-winner = \$0                & Returns 0 \\
      \rule{0pt}{0pt} \\
      \hline
    \end{tabular}
  \end{center}
\end{table}
\normalsize
Any unclaimed monies in the \emph{Pools Fund} are added to the current jackpot and carried over to the next drawing. From Table \ref{Ta:PrizeAllocIn6492013-}, it is clear that the majority contribution to the carryover is the $79.5\%$ that occurs when there is no 6/6 winner. But 5/6+ and 5/6- tickets also have low probabilities of occurring and when there are no 5/6+ or 5/6- winners, those shares of 6\% and 5\%, respectively, are added to the carrover pool for the next lottery.

The probabilities of these tickets occurring in an equiprobable lottery are denoted by $p_1$, $p_2$, etc. This notation is useful in the analytical expressions developed below.

\small
\begin{exmp}[Example of prize payouts under the current 9/8/2013 rules]\label{Exmp:ExampleOfPrizePayouts}

The carryover is \$30,000,000 and the crowd bets $10,000,000$ tickets, of which $1,000,000$ are assumed to be free plays, yielding a net cash contribution of $\$27,000,000$. Assuming the crowd chooses quick picks with probabilities of $1/t$, numbers for each ticket will have a binomial distribution. Random selections under the binomial are displayed in column 3 of Table \ref{Ta:ExamplePrizeAllocIn6492013-}, which gives the probability of this class of ticket winning. The first column is the winning ticket type, the second, the number of combinations, the fourth the total payouts to the crowd, the fifth the number of tickets held by the syndicate and the sixth, the total payouts to the syndicate. Using these numbers, we calculate the \emph{prize pool}, the fixed payouts to crowd and syndicate, and the pools fund as follows:

\vspace{2mm}
\scriptsize
\begin{itemize}
\item Prize Pool:  $\$27,580,579 = 0.40 * (13,983,816 * 3 \, + \, 0.90*30,000,000)$.
\item Crowd Fixed: $\$4,077,490 = 176,933*\$10 + 123,569*\$5 + 1,198,805*\$1.41$.
\item Syndicate Fixed: $\$5,696,520 = 246,820*\$10 + 172,200*\$5 + 1,678,950*\$1.41$.
\item Pools Fund:  $\$17,806,569 = \$27,580,579 - \$4,077,490 - \$5,696,520$.
\end{itemize}

\scriptsize
\begin{table}[ht]
  \begin{center}
    \caption{\small{Example of Payouts from a Sample 6/49 Pools Fund.}}
    \label{Ta:ExamplePrizeAllocIn6492013-}
    \scriptsize
    \begin{tabular} {lrrrrr}
      \rule[-4pt]{0pt}{10pt} \\
      \hline
      \rule[-4pt]{0pt}{10pt} \\
             &               & \# Crowd & Crowd  & \# Syndicate & Syndicate \\
      Type   &  Combinations &  Tickets & Payout &  Tickets     &  Payout   \\
      \rule{0pt}{2pt} \\
      \hline
      \rule{0pt}{0pt} \\
      $6/6$  &          $1$ &         0 &         \$0 &          1 & \$44,156,222 \\
      $5/6+$ &          $6$ &         6 &   \$534,135 &          6 &    \$534,135 \\
      $5/6-$ &        $252$ &       185 &   \$375,960 &        252 &    \$514,534 \\
      $4/6$  &     $13,545$ &     9,773 &   \$708,909 &     13,545 &    \$982,518 \\
      $3/6$  &    $246,820$ &   176,933 & \$1,769,330 &    246,820 &  \$2,468,200 \\
      $2/6+$ &    $172,200$ &   123,569 &   \$617,845 &    172,200 &    \$861,000 \\
      $2/6-$ &  $1,678,950$ & 1,198,805 &         \$0 &  1,678,950 &          \$0 \\
      \rule{0pt}{0pt} \\
      \hline
    \end{tabular}
  \end{center}
\end{table}

\small
Summing all payouts in the syndicate payout column gives \$49,516,609, for a gain of 
\[
  \$7,565,161 = \$49,516,609 - \$3*13,983,816.
\]
plus 1,678,200 free plays in the next lottery. The cash payout from non-6/6 tickets is \$5,360,387, despite a crowd and syndicate bet of \$68,951,454. Clearly, the jackpot must be large in order for buying the pot to be justifiable.

\end{exmp}
\normalsize

\end{description}

%************************************************************************************************
%************************************************************************************************
\section{Expected Return from Buying the Pot}\label{S:ExpectedReturnFromBuyingThePot}

%************************************************************************************************
\subsection{Notation and Terminology}

\normalsize
Table \ref{Ta:FixedParametersForThe649} gives the fixed parameters of the lottery, namely those that do not involve betting strategies of the syndicate or crowd. 

\begin{table}[ht]
  \footnotesize
  \begin{center}
    \caption{\footnotesize{Fixed Parameters for the 6/49 Lotto.}}
    \label{Ta:FixedParametersForThe649}
    \begin{tabular} {ll}
      \rule[-4pt]{0pt}{10pt} \\
      \hline
      \rule[-4pt]{0pt}{10pt} \\
      Notation & Description \\
      \rule{0pt}{2pt} \\
      \hline
      \rule{0pt}{0pt} \\
      $t$           & Number of tickets in the lottery = 13,983,816. \\
      $a$           & Carryover pool in dollars, $a \ge 0$. \\
      $p_i$         & Probability that a ticket is of class $i$ assuming that \\
                    & the winning ticket is drawn equiprobably (see Table \ref{Ta:PrizeAllocIn6492013-}). \\
      $f$           & Fraction of tickets that are ``free plays.'' \\
      $c$           & Number of tickets bet by the crowd. \\
      \rule{0pt}{0pt} \\
      \hline
    \end{tabular}
  \end{center}
\end{table}
\normalsize

Table \ref{Ta:RandomVariablesThatInvolveBettorStrategies} has the notation for the random variables that account for stochasticity and strategy in playing the lottery. 

\begin{table}[ht]
  \footnotesize
  \begin{center}
    \caption{\footnotesize{Random Variables for 6/49 Payouts.}}
    \label{Ta:RandomVariablesThatInvolveBettorStrategies}
    \begin{tabular} {ll}
      \rule[-4pt]{0pt}{10pt} \\
      \hline
      \rule[-4pt]{0pt}{10pt} \\
      Notation & Description \\
      \rule{0pt}{2pt} \\
      \hline
      \rule{0pt}{0pt} \\
      $N_i$         & Random variable for the number of tickets of class $i$ bet by\\
                    & \quad the crowd (see Table \ref{Ta:PrizeAllocIn6492013-}). \\
      $N$           & Vector $N = (N_1, N_2, \ldots, N_8)'$. \\
      $d_{\!_{AB}}$ & Dollars awarded or deducted for tickets of types (A) and (B). \\
      $d_{\!_{BP}}$ & Dollars in the \emph{betting pool}. \\
      $d_{\!_{PP}}$ & Dollars in the \emph{prize pool}. \\
      $d_{\!_{PF}}$ & Dollars in the \emph{Pools Fund}. \\
      \rule{0pt}{0pt} \\
      \hline
    \end{tabular}
  \end{center}
\end{table}
\normalsize
 
Using the notation in Tables \ref{Ta:FixedParametersForThe649} and \ref{Ta:RandomVariablesThatInvolveBettorStrategies}, the number of dollars in each fund is
\begin{align}
  d_{\!_{AB}} &= 10 (N_5 + t p_5) \, + \, 5 (N_6 + t p_6) \, + \, 1.41 (N_7 + t p_7) \label{dAB} \\
  d_{\!_{BP}} &= 3 ( t + c (1 - f)) \label{dBP} \\
  d_{\!_{PP}} &= d_{\!_{BP}} 0.4 \label{dPP} \\
  d_{\!_{PF}} &= d_{\!_{PP}} - d_{\!_{AB}} \label{dPF}
\end{align}

Since $f$ is non-random, the second entry of the above table is the only non-stochastic entry.

%************************************************************************************************
\subsection{Equiprobable Betting by the Crowd}\label{S:EquiprobableBettingByTheCrowd}

We calculate first the expected return to a syndicate that buys the pot when the crowd chooses tickets independently and equiprobably. As we discuss in Section \ref{S:Non-equiprobableBettingByTheCrowd}, this is the crowd's optimal strategy, although they do not employ it in practice --- and the cost of this ``mistake'' is considerable.

%************************************************************************************************
\subsubsection{Syndicate's Expected Value for Equiprobable Crowd Betting}

In Appendix \ref{S:TheSyndicate'sExpectedValueWhenTheCrowdBetsEquiprobably}, we develop a formula for the syndicate's expected gain $G(c)$ from the wagering of $\$41,951,448 = \$3 * 13,983,816$ on $13,983,816$ tickets:
\begin{align}
  E[ \, G(c) \, ] & = (a + 0.795*\mu(c)) \, \lambda(c)^{-1} (1 - exp(-\lambda(c))) \label{E:SyndicateExpectedReturn} \\
                  & + \left( 0.06 \nu(c) \, + \, 0.145 \frac{1}{1 + c/t} \right) \mu(c) \nonumber \\
                  & + \$3,329,200 \, - \, \$3 t, \nonumber
\end{align}
where
\begin{align*}
  \lambda(c) &= c/t, \\
  \mu(c)     &= 0.40 \cdot 3 \left( ( t + c \cdot (1 - f)) \right) - \, (t + c) \cdot (p_5 \$10 \, + \, p_6 \$5 \, + \, p_7 \$1.41), \\
  \nu(c)     &= E \left[ \frac{6}{6 + X_{\text{5/6+}}} \right], \qquad \text{for } X \sim Bin(c,6/t).
\end{align*}
The term $\nu(c)$ is calculated using the recursive formula in Appendix \ref{S:TheSyndicate'sExpectedValueWhenTheCrowdBetsEquiprobably} and appears as the last column of Table \ref{Ta:BettingThresholds}.

%************************************************************************************************
\subsubsection{Parameters that Lead to Positive Expected Return}
\label{SSS:ParametersThatLeadToPositiveExpectedReturn}

Consider the implications of the 6/49 rules and of Formula \eqref{E:SyndicateExpectedReturn}. Because the lottery sponsors take such a high percentage of the betting pool (60\%), a large jackpot is needed for a syndicate to have a positive expected return. When a syndicate bets one of each ticket, previous analysis showed the syndicate's numbers of winning tickets are known exactly, irrespective of the winning numbers from the drawing. There will always be exactly 1 winning ticket, exactly 6 5/6+ tickets, exactly 252 5/6- tickets, and so on. The RHS of first line of formula \eqref{E:SyndicateExpectedReturn} dominates the others when a jackpot $a$ is large.

Table \ref{Ta:BettingThresholds} shows the results of applying formula \eqref{E:SyndicateExpectedReturn} for 10 levels of total crowd betting (\emph{c}) to solve for the sizes of carryover pools (\emph{a}) that produce expected returns of 0\%, 10\% and 20\% for the syndicate. Since the cost of buying the pot is $\$3 * 13,983,816 =\$41,951,448$, a return of 10\% is $\$4,195,145$. When the crowd bets \$30 million, for example, any carryover larger than \$36.92 million is a potential play for the syndicate, and carryovers of \$42.80 and \$48.67 million have expected returns of 10\% and 20\%, respectively. The last three columns of Table \ref{Ta:BettingThresholds} provide insight into the payout structure. The sixth column shows the expected amount in the \emph{Pools Fund} and the next column is its percentage in the \emph{prize pool}. Thus when the crowd bets \$40 million, the expected \emph{Pools Fund} is \$19.97 million, which is 49.68\% of the prize pool. Thus, the charges for fixed payout tickets amount to \$50.32\% of the \emph{prize pool}. The final column is the expected value for the 5/6+ factor:
\begin{equation}
  \text{EV56+} = E \left[ \frac{6}{6 + X_{\text{5/6+}}} \right]. \label{E:ExpValOf5/6+Factor} 
\end{equation}
The expected value declines when the crowd bets more, as one expects since $X_{\text{5/6+}}$ is generally larger.

\begin{table}[htp!]
  \begin{center}
    \caption{\small{Carryover thresholds from buying the pot for breakeven, 10\% and 20\% returns as a function of the size of the crowd's total bet, assuming the $\bm{f = 10\%}$ of the crowd's tickets are free plays. The sixth column has the expected pools fund, the seventh, the expected percentage of the pools fund to the prize pool and the last, the expected value of the 5/6+ factor of expression \eqref{E:ExpValOf5/6+Factor}.}} 
    \label{Ta:BettingThresholds}
    \scriptsize
    \begin{tabular} {rccccccc}
      \rule[-4pt]{0pt}{10pt} \\
      \hline
      \rule[-4pt]{0pt}{10pt} \\
        \multicolumn{1}{c}{Crowd}      &  Crowd     &  \multicolumn{3}{c}{Carryover Thresholds} & Expected   & (Pools Fund) \textdiv & \text{EV56+}  \\
        \multicolumn{1}{c}{Tickets}    &  \$ Bet    &  Breakeven  & +10\%       & +20\%         & Pools Fund & (Prize Pool)          & Eqn. \eqref{E:ExpValOf5/6+Factor} \\
        \multicolumn{1}{c}{(millions)} & (millions) & (millions)  & (millions)  & (millions)    & (millions) &  (\%)                 &  (\%) \\
      \rule{0pt}{2pt} \\
      \hline
      \rule{0pt}{4pt} \\
              3.3 & 9  & 30.33 & 35.05 & 39.76 & 13.28 & 58.39 & 82.71 \\
              6.7 & 18 & 33.46 & 38.74 & 44.01 & 15.51 & 54.02 & 70.15 \\
             10.0 & 27 & 36.92 & 42.80 & 48.67 & 17.74 & 51.32 & 60.71 \\
             13.3 & 36 & 40.71 & 47.22 & 53.73 & 19.97 & 49.68 & 53.40 \\
             16.7 & 45 & 44.81 & 51.99 & 59.17 & 22.20 & 48.73 & 47.60 \\
             20.0 & 54 & 49.21 & 57.10 & 64.99 & 24.44 & 48.27 & 42.90 \\
             23.3 & 63 & 53.90 & 62.52 & 71.15 & 26.67 & 48.14 & 39.02 \\
             26.7 & 72 & 58.84 & 68.24 & 77.63 & 28.90 & 48.25 & 35.77 \\
             30.0 & 81 & 64.03 & 74.23 & 84.42 & 31.13 & 48.53 & 33.01 \\
             33.3 & 90 & 69.45 & 80.46 & 91.48 & 33.37 & 48.92 & 30.64 \\
      \rule{0pt}{4pt} \\
      \hline
    \end{tabular}
  \end{center}
\end{table}
\normalsize

Recall from Example \ref{Exmp:ExampleOfPrizePayouts} that the crowd bet a net \$27,000,000 on 10 million tickets and the carryover was \$30,000,000 --- yet the syndicate won over \$6 million. According to Table \ref{Ta:BettingThresholds}, the syndicate should not bet under these conditions, since a minimum carryover of \$36.92 million is necessary. There is no problem here, since the numbers in the table are expected values and it is quite possible for a syndicate to win despite making an unfavorable bet. The syndicate in that example just got lucky. 

%************************************************************************************************
\subsection{Non-equiprobable Betting by the Crowd}\label{S:Non-equiprobableBettingByTheCrowd}

The calculations in Section \ref{S:EquiprobableBettingByTheCrowd} assumed that the crowd bets independently using $q = \frac{1}{t} 1_t$, where $1_t$ is a t-vector of all ones. What happens when the crowd bets using $q \ne \frac{1}{t} 1_t$? 

In part \ref{Enum:OptimalStrategy}, we stated a result from \cite{moffitt:ziemba:2017a}, that for \emph{pure jackpot} lotteries (ones having a single prize, a non-stochastic jackpot\footnote{We are assuming that the crowd's number of tickets, $c$, is known.} $v$) the expected payoff is
\begin{equation}
  E_q \left[ v \frac{1}{1 + N_1} \right] \, = v E_q \left[ \frac{1}{1 + N_1} \right] \, > \, v E_{1_t/t} \left[ \frac{1}{1 + N_1} \right] = \, E_{1_t/t} \left[ v \frac{1}{1 + N_1} \right], \label{E:CrowOptimalqi}  
\end{equation} 
where $q \ne 1/t 1_t$, $N_1$ is the random number of 6/6 tickets held by the crowd. However, formula \eqref{E:CrowOptimalqi} does not apply in the present case because $v$ is stochastic, depending on the size of the \emph{Pools Fund}.

Consider a non-stochastic configuration of single ticket bets $n_j = (n_{j1}, n_{j2}, \ldots, n_{jt})'$ for individuals $j = 1, \ldots, c$, each having zeroes except for a single $1$ in some position. Define $z_k = \sum_{j=1}^{j=c} n_{jk}$ and t-vector $z = (z_1, \ldots, z_t)'$. Clearly, $\sum_{k=1}^{k=t} z_k = c$. To compute the expected values of ticket types 1, 2, \ldots 7 with respect to an equiprobable drawing, observe that as $i$ ranges over all ticket drawings $i = 1, \ldots t$, for \textbf{any} $n_j$, the number of 6/6 is 1, the number of 5/6+ is 6, the number of 5/6 is 252, and so on as indicated in Table \ref{Ta:PrizeAllocIn6492013-}. Since the drawing is equiprobable, dividing each of these by $t$ gives the probability that \textbf{any} non-stochastic ticket will be of the indicated type under an equiprobable drawing. Define indicator functions on single ticket t-vectors $n$ as:
\[
  I_{x/6}(n) = \begin{cases}
                 1 & \text{if ticket } n \text{ is a x/6 ticket}, \\
                 0 & \text{otherwise.} 
               \end{cases}
\]
Applying this to fixed payout types 3/6, 2/6+ and 2/6, we obtain for $d_{\!_{AB}}$ in formula \eqref{dAB} 
\begin{align}
  E_e[d_{\!_{AB}}] &= E_e \left[ \sum_{j=1}^{j=c} \left\{ \$10 I_{3/6}(n_j) \, + \, 5 I_{2/6+}(n_j) \, + \, 1.41 I_{2/6} \right\} \right] + \$5,696,520 \nonumber \\
                   &= \left( \$10 p_{3/6} \, + \$5 \, p_{2/6+} \, + \, \$1.41 p_{2/6} \right) c \,  + \, \$5,696,520, \nonumber \\
                   &= \$0.4073651 \cdot c * \, + \, \$5,696,520. \label{E:ExpectedFixedTicketPayout}
\end{align}
where the notation $E_e$ emphasizes that the expectation is taken over equiprobable drawings and $\$5,696,520$ is the fixed payout/deduction for the syndicate.\footnote{$\$5,696,520 = \$10*246,820 + \$5*172,200 + \$1.41*1,678,950$.} The (stochastic) jackpot is $v = a \, + \, 0.795 d_{\!_{PF}}$ and the random 6/6 payout to the syndicate is 
\begin{align}
  v \frac{1}{1 + N_1} &= \left( a \, + \, 0.795 d_{\!_{PF}} \right) \frac{1}{1 + N_1} \nonumber \\
                      &= \left( a \, + \, 0.795 (0.4 (3( t \, + \, c(1 - f))) - d_{\!_{AB}}) \right) \frac{1}{1 + N_1} \nonumber \\
                      &= \left( a \, + \, 0.954( t \, + \, c(1 - f)) \right) \frac{1}{1 + N_1} - \frac{d_{\!_{AB}}}{1 + N_1} \nonumber \\
                      &= \left( a \, + \, 0.954( t \, + \, c(1 - f)) - \$5,696,520 \right) \frac{1}{1 + N_1} - \label{E:SyndicateTermOfJackpotPayout} \\
                      & \qquad \frac{\$10 N_5 \, + \, 5 N_6 \, + \, 1.41 N_7}{1 + N_1} \label{E:CrowdTermOfJackpotPayout}
\end{align}
In \eqref{E:SyndicateTermOfJackpotPayout}, the factor multiplying $1/(1 + N_1)$ is fixed. Its expectation using \eqref{E:CrowOptimalqi} is
\begin{align}
 & E_q \left[ \left( a \, + \, 0.954( t \, + \, c(1 - f) - \$5,696,520) \right) \frac{1}{1 + N_1} \right] \nonumber \\
 &   \qquad > \left( a \, + \, 0.954( t \, + \, c(1 - f) - \$5,696,520) \right) \frac{1}{\lambda} ( 1 - \exp(-\lambda) ), \label{E:SyndicateCostBound}
\end{align}
where $\lambda = c/t$. Thus for this term at least, the syndicate gets more than a fair split of the jackpot since 
\[
  \frac{1}{\lambda} ( 1 - \exp(-\lambda) ) > \frac{t}{t + c}.
\]
The second term \eqref{E:CrowdTermOfJackpotPayout} depends on $N_1$, $N_5$, $N_6$ and $N_7$, which respectively, are the numbers of 6/6, 3/6, 2/6+ and 2/6 tickets held by the crowd, and these are dependent on the crowd betting probabilities $q = (q_1, \ldots, q_t)'$. But we do not have the data to model the joint distribution of $(N_1, N_5, N_6, N_7)$ which is needed to evaluate \eqref{E:CrowdTermOfJackpotPayout}. 

However, we have circumstantial evidence that $N_5$, $N_6$ and $N_7$ are positively correlated with $N_1$. Therefore we make a crude assumption that the joint crowd payouts for 3/6, 2/6+ and 2/6 tickets are increased linearly with the winning ticket, that is, the payout for ticket $i$ is proportional to $q_i$:
\[
  \frac{\$10 N_5 \, + \, \$5 N_6 \, + \, \$1.41 N_7}{1 + N_1} \cdot q_i / (1/t).
\]
Thus if the winning ticket $i$ is bet with twice the frequency of an equiprobable bet, so that $t q_i = 2$, then the fixed payouts/deductions will be twice that expected in the equiprobable case (see the discussion leading to equation \eqref{E:ExpectedFixedTicketPayout}).

Using $H = \$10 p_5 \, + \, \$5 p_6 \, + \, \$1.41 p_7$, we calculate
\begin{align}
  E_{1_t/t} \left[ \min \left( c H t q_i, d_{PF} \right) \frac{1}{1 + N_1} \right] 
      &\le E_{1_t/t} \left[ c H t q_i \frac{1}{1 + N_1} \right] \nonumber \\
      &=  H t E_{1_t/t} \left[ c q_i \frac{1}{c q_i} (1 - e^{-cq_i}) \right] \nonumber \\
      &=  H t \sum_{i=1}^{i=t} \frac{1}{t} (1 - e^{-cq_i}) \nonumber \\
      &\le H t (1 - e^{-c/t}) \label{E:CrowdCost} \\
      &= \frac{c H}{\lambda} (1 - e^{-\lambda}), \label{E:CrowdCostBound}
\end{align}
where $\lambda = c/t$ and the step \eqref{E:CrowdCost} follows from Jensen's inequality since $1 - e^{-cq}$ is a concave function of $q$. Jensen's inequality can be stated as follows. A function $f: [a,b] \rightarrow \mathbb{R}$ that satisfies $f(ta + (1-t)b) \le t f(a) + (1-t) f(b)$ for all $t \in (0,1)$ is called \emph{convex}, and if the inequality is strict, \emph{strictly convex}. For a random variable $X$ and \emph{convex} function $f$, Jensen's inequality asserts that $f(E[X]) \le E[f(X)]$. Further, if $X$ is not degenerate and $f$ is strictly convex, then $f(E[X]) < E[f(X)]$. A function $f$ is (\emph{strictly}) \emph{concave} if $-f$ is (\emph{strictly}) \emph{convex}i, so Jensen's inequality is reversed for \emph{concave} functions.

Putting \eqref{E:SyndicateCostBound} together with \eqref{E:CrowdCostBound} yields
\begin{align}
   E \left[v \frac{1}{1 + N_1} \right] &\ge \left( a \, + \, 0.954( t \, + \, c(1 - f)) - \$5,696,520 \, - \, c H \right) \nonumber \\
                                       &\qquad \cdot \frac{1}{\lambda} (1 - e^{-\lambda}) \label{E:SyndicateExpectedReturn}
\end{align}
where $\lambda = c/t$ and $H = \$10 p_5 \, + \, \$5 p_6 \, + \, \$1.41 p_7$.

This calculation shows that the syndicate obtains a better result than when the crowd bets proportionally, as in the corresponding result for \emph{pure jackpot} lotteries.

%************************************************************************************************
%************************************************************************************************
\section{Design Considerations for Lotteries}\label{SS:DesignConsiderationsForLotteries}

Lottery design includes the goal to maximize the the sponsors' earnings. Assuming fairly constant fixed costs of running the lottery, sponsors should strive to make the lottery popular, thereby increasing profitability. The most recent changes to payouts were made with that goal in mind --- these changes increased the ``convexity'' of payouts, meaning many little prizes and greater jackpot growth. Ziemba recommended these designs in his work in the 1980's and \cite{Walker2008459} later also recommended them. Convex designs encourage players because more ``get something back,'' while at the same time growing large jackpots quickly. 

This design is supported by research in behavioral finance. Lopes' SP/A (Security-Potential/Aspiration) model (\cite{Lopes1987255}), an improved version of the classic Friedman/Savage (1948) utility curves, argues that many unsophisticated gamblers prefer strategies of buying safe prospects with a few longshots (the ``Cautiously Hopeful'' pattern of SP/A). Regarding large \emph{jackpots}, Daniel Kahneman has written

\small
\begin{quotation}
  \noindent ``For emotionally significant events, the size of the 
  probability simply doesn't matter. What matters is the possibility 
  of winning. People are excited by the image in their mind. The 
  excitement grows with the size of the prize, but it doesn't diminish 
  with the size of the probability.''  Source: 
  \cite{NYTimes:YourMoney:KahnemanQuote:Online}.
\end{quotation}
\normalsize 

There is another aspect of lottery design, namely, discouraging syndicates from buying the pot. There are two ways to accomplish this: (1) creating a large number of tickets making it logistically difficult to buy the pot, and (2) using convex designs, which reduces the likelihood that pot buying situations will occur. 
Method (1) is not feasible except for large lotteries like the California Powerball lottery. The reason is that if the number of tickets sold are too small relative to the total number of tickets, the jackpot may build slowly and seldom be won. On the other hand, method (2) can be effective regardless of the size of the lottery. To illustrate, consider a \emph{pure jackpot} lottery with the same carryover, take and crowd betting as in Table \ref{Ta:BettingThresholds}. The results are shown in Table \ref{Ta:BettingThresholdsForPureJockpotAnd649}. The first column has the number of tickets, which after a 10\% deduction for free plays, equals the crowd contribution to the \emph{betting pool} shown in the second column. Then assuming a take of 60\%, breakeven thresholds of 0\%, 10\% and 20\% for the pure jackpot lottery are shown in columns 3-5 and for the 6/49 in columns 6-8. The results show that buying the pot thresholds are lower in the pure lottery, but not as much as might be expected.

But one can see the reason by a simple argument. When the sponsors takes 60\%, only 40 cents is returned as prizes for each dollar wagered. Therefore, a syndicate needs to recover 60\% of the covering bet, or $0.6*\$3*13,983,816 = \$25,170,869$, regardless of the lottery's rules. As we have shown, the syndicate earns its fair share of consolation prizes, but the free plays it earns are not worth too much since after the lottery is hit the next lottery when those tickets will be used will have a small purse. Assuming the the crowd bets $\$1,000,000$ on the next lottery the expected value of these $1,678,950$ tickets will under optimal wagering be worth about $\$150,000$.
\begin{table}[htp!]
  \begin{center}
    \caption{\small{Carryover thresholds for a \emph{pure jackpot} lottery and the 6/49 Lotto.}} 
    \label{Ta:BettingThresholdsForPureJockpotAnd649}
    \scriptsize
    \begin{tabular} {rccccccc}
      \rule[-4pt]{0pt}{10pt} \\
      \hline
      \rule[-4pt]{0pt}{10pt} \\
        \multicolumn{1}{c}{Crowd}      &  Crowd     &  \multicolumn{3}{c}{Carryover Thresholds for Pure Jackpot} & \multicolumn{3}{c}{Carryover Thresholds for 6/49 Lotto} \\
        \multicolumn{1}{c}{Tickets}    &  \$ Bet    &  Breakeven  & +10\%       & +20\%         &  Breakeven  & +10\%       & +20\%      \\
        \multicolumn{1}{c}{(millions)} & (millions) & (millions)  & (millions)  & (millions)    & (millions)  & (millions)  & (millions) \\
      \rule{0pt}{2pt} \\
      \hline
      \rule{0pt}{4pt} \\
              3.3 & 9  & 26.77 & 31.48 & 36.20 & 30.33 & 35.05 & 39.76 \\
              6.7 & 18 & 28.76 & 34.04 & 39.31 & 33.46 & 38.74 & 44.01 \\
             10.0 & 27 & 31.14 & 37.02 & 42.89 & 36.92 & 42.80 & 48.67 \\
             13.3 & 36 & 33.90 & 40.41 & 46.92 & 40.71 & 47.22 & 53.73 \\
             16.7 & 45 & 37.02 & 44.20 & 51.38 & 44.81 & 51.99 & 59.17 \\
             20.0 & 54 & 40.49 & 48.38 & 56.26 & 49.21 & 57.10 & 64.99 \\
             23.3 & 63 & 44.28 & 52.91 & 61.53 & 53.90 & 62.52 & 71.15 \\
             26.7 & 72 & 48.37 & 57.77 & 67.17 & 58.84 & 68.24 & 77.63 \\
             30.0 & 81 & 52.75 & 62.94 & 73.13 & 64.03 & 74.23 & 84.42 \\
             33.3 & 90 & 57.38 & 68.39 & 79.41 & 69.45 & 80.46 & 91.48 \\
      \rule{0pt}{4pt} \\
      \hline
    \end{tabular}
  \end{center}
\end{table}
\normalsize

We conclude the discussion by examining the impacts of design choices in the 6/49 Lotto. The 6/49 Lotto's convex design according to Table \ref{Ta:BettingThresholdsForPureJockpotAnd649} raised the bar for buying-the-pot strategies, making carryover thresholds roughly 12\% to 20\% higher. We now compare the impacts of the 6/49's design features toward increasing the threshold for buying the pot. We identify four factors: (1) the take, (2) the payouts for small prizes, (3) the payouts for large, non 6/6 prizes, and (4) free plays. Then we compare by
\begin{enumerate}
\item Changing the take only, using alternatives 55\%, 60\% (current) and 65\%.
\item Eliminating fixed payouts 3/6 and 2/6+ only.
\item Eliminating 4/6, 5/6 and 5/6+ payouts only.
\item Eliminating free plays only.
\end{enumerate}
Table \ref{Ta:BettingThresholdsForDesignFactors} shows breakeven carryover thresholds for these design factors. The factor is indicated in the first column and the other 5 columns are carryover thresholds when the crowd bets the indicated millions of dollars, 20, 40, etc. In the second column (corresponding to a crowd bet of \$20 million), the numbers in parenthesis are differences of threshold carryovers from the current 6/49 values (second row, second column). Since the relative impacts of these factors are the same for the five crowd betting amounts, their impacts on the buying the pot strategy can be assessed using this column. The greatest factor impact is due to free plays; removing them drops the threshold by $\$3.39$ million ($\sim 10\%$). The largest inhibitor is clearly the take --- increasing it by $0.05\%$ from to $65\%$ has a large impact on breakeven carryovers. 

\begin{table}[htp!]
  \begin{center}
    \caption{\small{Breakeven Carryover Thresholds for Various 6/49 Design Factors.}} 
    \label{Ta:BettingThresholdsForDesignFactors}
    \scriptsize
    \begin{tabular} {lcccccccccc}
      \rule[-4pt]{0pt}{10pt} \\
      \hline
      \rule[-4pt]{0pt}{10pt} \\
        \multicolumn{1}{c}{      } &  \multicolumn{5}{c}{Crowd Bets in Millions of Dollars} \\
        \multicolumn{1}{c}{Design} &   20     &   40    &   60    &   80    &   100    \\
        \multicolumn{1}{c}{Factor} & million & million  & million & million & million  \\
      \rule{0pt}{2pt} \\
      \hline
      \rule{0pt}{4pt} \\
        TAKE=0.55     & 30.56 (-2.90) & 36.98 & 44.64 & 53.41 & 63.15 \\
        CURRENT 6/49  & 33.46 ( 0.00) & 40.71 & 49.21 & 58.84 & 69.45 \\
        TAKE=0.65     & 36.37 ( 2.91) & 44.44 & 53.79 & 64.27 & 75.74 \\
        NO 2/6+, 3/6  & 32.88 (-0.58) & 39.65 & 47.76 & 57.08 & 67.44 \\
        NO 4/6, 5/6   & 32.99 (-0.47) & 39.87 & 48.07 & 57.48 & 67.91 \\
        NO FREE PLAY  & 30.07 (-3.39) & 36.28 & 43.73 & 52.29 & 61.81 \\
      \rule{0pt}{4pt} \\
      \hline
    \end{tabular}
  \end{center}
\end{table}
\normalsize

Based on these statistics, we make recommendations for state lotteries using ratings of the form $(+=-\pm\mp, +=-\pm\mp)$. The first sign is for popularity, the second for inhibiting buyers of the pot. For example, $(+,-)$ indicates that a factor increases the lottery's popularity, but encourages buying the pot.
\begin{enumerate}
\item $(\mp,+)$ If possible, add combinations to the lottery by increasing the numbers.
\item $(+,+)$ Initiate a free play feature.
\item $(\mp,+)$ Increase the take.
\item $(+,=)$ Offer many small prizes.
\item $(+,=)$ Increase the allocation of the \emph{Pools Fund} to 6/6 winners.
\item $(=,+)$ Decrease the awards to hard-to-win non 6/6 tickets.
\end{enumerate}
Increasing the allocation to 6/6 allows quicker build-up of jackpots, which encourages greater crowd participation. However, we did not address the question of build-up speed of the jackpot, nor the acceleration of betting on larger jackpots. These need to be studied in order to design prizes and allocations to optimize betting flows. 

%*********************************************************************************************
%*********************************************************************************************

\section{Conclusions}

In this paper, we have shown conditions under which buying the pot in the 6/49 Lotto has positive expected return when the crowd bets equiprobably. We also indicated that equiprobable betting is optimal for the crowd, that is, expected return is lower when it does not bet equiprobably. We illustrated the advantages of lotteries with convex designs by calculating 6/49 carryover thresholds and comparable \emph{pure jackpot} carryover thresholds. We then rated various design features for their likelihood of increasing a lottery's popular, and decreasing the likelihood of buyers of the pot.

%*********************************************************************************************
%*********************************************************************************************
\begin{appendices}

%************************************************************************************************

\section{The Syndicate's Expected Value when the Crowd bets Equiprobably}\label{S:TheSyndicate'sExpectedValueWhenTheCrowdBetsEquiprobably}

Assuming that the lottery's tickets are equiprobable, $(N_1, \ldots N_8)'$ has a multinomial distribution 
\begin{equation}
  (N_1, \ldots N_8)' \sim Multin(c + t, p),
\end{equation}
where $p = (p_1, p_2, \ldots, p_8)'$. The distribution of deductions $d_{\!_{AB}}$ from the \emph{prize pool} is given by 
\begin{equation}
  d_{\!_{AB}} = \beta_{\!_{AB}}' N,
\end{equation}
where $\beta_{\!_{AB}} = (0,0,0,0,10,5,1.41,0)'$ and $N \sim Multin(c + t, p)$. 

Substituting \eqref{dBP} and \eqref{dPP} into equation \eqref{dPF} gives the \emph{Pools Fund} as
\begin{equation}
  d_{\!_{PF}} = 0.40 \cdot 3 \cdot ( t + c \cdot (1 - f)) \, - \, d_{\!_{AB}}. 
\end{equation}
and in RHS of this expression, only $d_{\!_{AB}}$ is random.

Using results from \cite{moffitt:ziemba:2017a}, the expected value $G$ of the syndicate's net gain, given $d_{\!_{PF}}$, as
\begin{align}
  E[ G(c) \, | \, d_{\!_{PF}}] &= 0.795 \cdot (a + d_{\!_{PF}}) E \left[ \frac{1}{1 + X_{\text{6/6}}} \right] \label{E:6/6Expectation} \\
                               & + 0.06 \cdot d_{\!_{PF}} E \left[ \frac{6}{6 + X_{\text{5/6+}}} \right] \label{E:5/6+Expectation} \\
                               & + 0.05 \cdot d_{\!_{PF}} E \left[ \frac{252}{252 + X_{\text{5/6-}}} \right] \label{E:5/6-Expectation} \\
                               & + 0.095 \cdot d_{\!_{PF}} E \left[ \frac{13545}{13545 + X_{\text{4/6}}} \right] \label{E:4/6Expectation} \\
                               & + \$2,468,200 \, + \, \$86,100 \, - \, \$3 t \label{E:2-3/6Expectation}
\end{align}
where 
\begin{align*}
  (X_{\text{6/6}}, X_{\text{5/6+}}, X_{\text{5/6-}}, X_{\text{4/6}})' \sim Multin(c,(1, 6, 252, 13545)'/t).
\end{align*}
Since buying one of each ticket gives the same exact payout regardless of the winning ticket (numbers of tickets shown in Table \ref{Ta:PrizeAllocIn6492013-}), we know that a covering strategy pays \$2,468,200 and \$86,100, respectively, for 3/6 and and 2/6+ tickets. This explains the term \eqref{E:2-3/6Expectation}.

\noindent Using the formulas from \eqref{E:EVrecursion} we obtain for $\lambda(c) = c/t$
\begin{equation}
  E \left[ \frac{1}{1 + X_{\text{6/6}}} \right] = \lambda(c)^{-1} (1 - exp(-\lambda(c)))
\end{equation}
and values $\nu(c) = E \left[ \frac{6}{6 + X_{\text{5/6+}}} \right]$ using recursion. These calculations take care of terms \eqref{E:6/6Expectation} and \eqref{E:5/6+Expectation}.

Using the Law of Large Numbers, the expectation in the term \eqref{E:5/6-Expectation} can be approximated by 
\begin{equation}
  \frac{252}{252 \, + \, 252 c/t} = \frac{1}{1 + c/t}. \label{E:FairSplitExpectation3}
\end{equation}
and in term \eqref{E:4/6Expectation} by 
\begin{equation}
  \frac{13545}{13545 \, + \, 13545 c/t} = \frac{1}{1 + c/t}. \label{E:FairSplitExpectation4}
\end{equation}
Basically, these two approximations amount to fair split of the corresponding share of the \emph{Funds Pool}. Thus 
\begin{align}
  E[ G(c) \, | \, d_{\!_{PF}}] & = (a + 0.795 \cdot d_{\!_{PF}}) \lambda(c)^{-1} (1 - exp(-\lambda(c))) \label{E:6/6ExpectationC} \\
                               & + 0.06 \cdot d_{\!_{PF}} E \left[ \frac{6}{6 + X_{\text{5/6+}}} \right] \label{E:5/6+ExpectationC} \\
                               & + 0.145 \cdot d_{\!_{PF}} \frac{1}{1 + c/t} \label{E:5/6-ExpectationC} \\
                               & + 2,468,200 \, + \, 86,100 \, - \, \$3 t \label{E:2-3/6ExpectationC} 
\end{align}
where $\lambda(c) = c/t$. To complete the calculation, we need to eliminate the dependence of $E[ G(c) \, | \, d_{\!_{PF}}]$ on $d_{\!_{PF}}$ by determining its distribution and performing an integration. But this is straightforward: the first three terms \eqref{E:6/6ExpectationC}, \eqref{E:5/6+ExpectationC} and \eqref{E:5/6-ExpectationC} are linear in $d_{\!_{PF}}$ so that the expectation $\mu(c) = E[d_{\!_{PF}}]$ should be substituted for $d_{\!_{PF}}$. The expectation $E[d_{\!_{PF}}]$can be calculated by substituting \eqref{dBP} into \eqref{dPP} and \eqref{dPP} into \eqref{dPF} taking expectations:
\[
   \mu(c) = E[d_{\!_{PF}}] =  0.40 \cdot 3 \cdot ( t + c \cdot (1 - f)) \, - \, E[d_{\!_{AB}}] 
\]
We calculate $E[d_{\!_{AB}}]$ as follows. For any ticket $i$, the number of tickets that are 3/6, 2/6+ and 2/6- are respectively $248,820$, $172,200$, and $1,678,950$, respectively. Therefore, the probability that ticket $i$ is a 3/6, 2/6+ or 2/6- ticket, \emph{given that a winning ticket is drawn equiprobably}, is $p_5 = 248,820/t$, $p_6 = 172,200/t$ and $p_7 = 1,678,950/t$, respectively. Now consider any choice of $c$ tickets. By linearity of expectations, the expected number of 3/6 tickets is $c * p_5$, of 2/6+ tickets is $c * p_6$ and of 2/6- tickets, $c * p_7$. Therefore,
\[
  \nu(c) = E[d_{\!_{AB}}] = (t + c) (p_5*\$10 + p_6*\$5 + p_7*\$1.41).
\]

Summarizing, the expected gain $G(c)$ to a syndicate that covers the pool is
\begin{align}
  E[ \, G(c) \, ] & = (a + 0.795 \cdot \mu(c)) \lambda(c)^{-1} (1 - exp(-\lambda(c)))  \label{E:SyndicateExpectedReturnAppendix}  \\
                  & + \, \left( 0.06 \nu(c) \, + \, 0.145 \frac{1}{1 + c/t} \right) \mu(c) \nonumber \\
                  & + \, \$2,553,300 \, - \, \$3 t, \nonumber
\end{align}
where
\begin{align*}
  \lambda(c) &= c/t, \\
  \mu(c)     &= 0.40 \cdot 3 \left( ( t + c \cdot (1 - f)) \right) - \, (t + c) \cdot (p_5 \$10 \, + \, p_6 \$5 \, + \, p_7 \$1.41), \\
  \nu(c)     &= E \left[ \frac{6}{6 + X_{\text{5/6+}}} \right], \qquad \text{for } X \sim Bin(c,6/t).
\end{align*}

\end{appendices}

\bibliographystyle{apalike}
\bibliography{DIPTBTP}

\end{document}